\documentclass[a4paper,11pt]{article}

\usepackage{jheppub} 

\usepackage[T1]{fontenc} 
\usepackage{amsmath,amssymb,bm}
\usepackage{hyperref}
\usepackage{mathtools}
\usepackage{physics}
\usepackage{graphicx}

\title{\boldmath A Composite Three-Body-Holonomy Toy Framework\\ for
Dynamical Electroweak-Scale Selection}

\author{Tadashi Yoshikawa,\note{Corresponding author.}}
\affiliation{Nagoya Aoi University \\Nagoya, Japan}
\emailAdd{tyoshika@nagoya-aoi.ac.jp}

\abstract{
We propose a phenomenological toy framework promoted to a minimal 
three-state internal sector in $(3+1)$ dimensions and
realized by composite strong dynamics. It is in a genuine three-body
holonomy which originates in a relativistic $(1+1)$-dimensional Dirac system.
The hypothesis is that the number three is not imposed first 
as a gauge symmetry, but is inherited from
the primitive closed cyclic structure of the three-body holonomy.  
This threefold structure subsequently reappears as a three-state internal space, the
three-site loop, a three-flavor hidden
sector, and a cubic determinant interaction.

We introduce an asymptotically free hidden $SU(3)_H$ gauge theory 
with three Standard-Model-singlet Dirac flavors. Strong hidden dynamics generates 
a composite condensate, while the inherited holonomy selects 
its cyclic internal orientation. The anomalous cubic determinant interaction 
produces a holonomy-dependent potential, and dimensional transmutation generates 
the characteristic composite scale.

The radial composite excitation along the cyclic direction, denoted by $S_3$, 
couples to the Standard-Model Higgs doublet through a holonomy-dependent 
CP-even portal. The resulting condensate shifts the Higgs quadratic term 
and can trigger electroweak symmetry breaking. 
In the classically scale-invariant reference limit, the electroweak scale 
is induced from the dynamically generated hidden-sector scale 
through dimensionless portal couplings.

The framework is not intended as a complete ultraviolet solution 
to the hierarchy problem, since an elementary Higgs remains ultraviolet sensitive. 
Rather, it provides a proof of concept for electroweak-scale selection 
through three-body holonomy and dimensional transmutation. 
A representative one-loop Coleman--Weinberg analysis indicates 
that an electroweak-scale minimum can persist in the reference parameter region. 
The mapping from the $(1+1)$-dimensional holonomy to the $(3+1)$-dimensional 
internal sector remains an explicit modeling assumption.
}

\begin{document} 
\maketitle
\flushbottom

\section{Introduction}

The large separation between the Planck scale and the electroweak scale,
\begin{equation}
 \frac{M_{\rm EW}}{M_{\rm Pl}}\sim 10^{-17},
\end{equation}
remains one of the central structural puzzles of particle physics.
Conventional approaches attempt to protect or generate the weak scale through
symmetry, compositeness, extra-dimensional geometry, or radiative dynamics.
A recurring lesson is that exponentially separated scales can emerge from
moderate dimensionless couplings through dimensional transmutation and
radiative dynamics rather than from an equally tiny fundamental mass ratio
\cite{ColemanWeinberg1973,GildenerWeinberg1976,Farzinnia2013,Chway2014}.
More generally, interactions between a hidden scalar sector and the
Standard-Model Higgs through operators of the form
$S^2H^\dagger H$ constitute the familiar Higgs-portal framework
\cite{Lebedev2021}.

The present work explores a toy mechanism motivated by genuine three-body
physics.  The starting point is a relativistic three-body Dirac system in
$(1+1)$ dimensions with pairwise contact interactions of the
Sakamoto--Munakata--Ino type \cite{Munakata1990,Sakamoto1993,SakamotoNakanoYoshikawa1993}.  
In that system
the triple-coincidence structure admits a genuine three-body holonomy
generated by
\begin{equation}
 Q_3=\alpha_1\alpha_2\alpha_3,
\end{equation}
and the corresponding phase can affect the invariant-mass spectrum
\cite{Yoshikawa2026phase,Yoshikawa2026family}.  A related consequence of the
same three-state cyclic structure is that, at the enhanced-symmetry point
$\theta_3=0$, the resulting $3\to2+1$ splitting is proportional to the
Cartan generator $T_8$ and selects the conventional
$SU(3)\to SU(2)\times U(1)$ direction \cite{Yoshikawa2026SU3}.

A central question of the present work is why the number three should persist
when the construction is lifted to an internal sector in $(3+1)$ dimensions.
Our working answer is minimality.  Two-body data can define links and matching
conditions, but the primitive genuine many-body cyclic structure considered
here first closes for three.  Four and higher site constructions are possible,
but they are nonminimal extensions of the original three-body source.  We
therefore use the phrase ``nature chooses three'' in the restricted sense that
three is the smallest nontrivial structure capable of faithfully carrying the
genuine many-body holonomy inherited from the starting system.  The recurring
pattern to be tested is
\begin{equation}
 3_{\rm body}
 \longrightarrow
 3_{\rm cyclic}
 \longrightarrow
 3_{\rm site}
 \longrightarrow
 3_{\rm flavor}
 \longrightarrow
 \det\Sigma\sim\Sigma^3.
 \label{eq:three_selection_intro}
\end{equation}
The hidden gauge theory introduced below is therefore not the origin of the
number three; it is a concrete dynamical realization of a threefold structure
selected earlier by the holonomy.

Our purpose is not to transplant the $(1+1)$-dimensional configuration-space
topology literally into four dimensions.  Instead, we retain its cyclic
information as a finite-dimensional internal degree of freedom and ask whether
a conventional four-dimensional strongly coupled sector can realize both the
three-site holonomy and a physical dynamical scale.  The scale hierarchy is
organized as
\begin{equation}
 M_{\rm Pl}
 >
 \Lambda_{\rm match}
 >
 \Lambda_{\rm 3B}\sim\Lambda_H
 >
 M_{\rm EW}.
 \label{eq:scale_chain_intro}
\end{equation}
Here $\Lambda_{\rm match}$ denotes an ultraviolet matching/onset scale for the
three-body effective description, while $\Lambda_{\rm 3B}$ denotes the
infrared scale dynamically generated by the composite three-body sector.  The
two scales need not coincide.

We use a QCD-like hidden $SU(3)_H$
theory with three Standard-Model-singlet Dirac flavors.  Its composite matrix
$\Sigma_{ij}\sim\bar\chi_{Rj}\chi_{Li}$ is particularly useful because for
three flavors the determinant interaction is cubic and contains the oriented
product $\Sigma_{12}\Sigma_{23}\Sigma_{31}$.  This provides a direct bridge
between a standard strong-interaction effective operator and the three-site
holonomy.  The strong dynamics then determines the magnitude of the condensate,
while the inherited three-body holonomy fixes its cyclic orientation.

It does not derive the
hidden gauge theory uniquely from the low-dimensional model and it does not
claim a complete solution of the ultraviolet hierarchy problem.  Rather, it
demonstrates that a three-body topological input, composite strong dynamics,
dimensional transmutation, and a holonomy-dependent Higgs portal can coexist
in one self-consistent four-dimensional toy EFT and can dynamically select an
electroweak-scale vacuum without inserting an elementary mass ratio of order
$10^{-17}$ by hand.

\section{Three-body holonomy in the $(1+1)$-dimensional Dirac model}

We consider three Dirac particles with a reference Hamiltonian
\begin{equation}
 H_0=
 \sum_{i=1}^{3}\alpha_i p_i
 +
 m\sum_{i=1}^{3}\beta_i,
\end{equation}
supplemented by pairwise contact interactions.
The pairwise Sakamoto--Munakata--Ino interactions can be transferred into
sector-dependent matching conditions, while the triple-coincidence structure
permits an additional global three-body phase
\cite{Munakata1990,Sakamoto1993,Yoshikawa2026phase}.

We introduce
\begin{equation}
 Q_3=\alpha_1\alpha_2\alpha_3,\qquad Q_3^2=1.
\end{equation}
The three-body holonomy is
\begin{equation}
 W_3=e^{i\theta_3 Q_3}
 =
 \cos\theta_3+iQ_3\sin\theta_3.
 \label{eq:W3}
\end{equation}
The kinetic and pair-interaction terms commute with $Q_3$, whereas the
constituent-mass operator anticommutes with it,
\begin{equation}
 [Q_3,K]=0,\qquad \{Q_3,M\}=0.
\end{equation}
Consequently, in the massless limit the Hilbert space separates into
$Q_3=q=\pm1$ sectors,
\begin{equation}
 W_3\ket{q}=e^{iq\theta_3}\ket{q},
\end{equation}
while a nonzero constituent mass mixes the two sectors.

At fixed intrinsic parity, the six ordering sectors may be reduced to a
three-dimensional cyclic space. A minimal $C_3$-invariant effective mass
operator then gives
\begin{equation}
 M^{[k]}
 =
 M_0+
 2\Delta
 \cos\left(
 \frac{\theta_3+2\pi k}{3}
 \right),
 \qquad k=0,1,2,
\end{equation}
showing explicitly how three-body cyclicity and holonomy can generate a
triplicated spectral structure \cite{Yoshikawa2026family}.

\section{Matching scale and dynamical three-body scale}
\label{sec:matching_scale}

We distinguish two scales that were conflated in simpler versions of the
construction.  The first is an ultraviolet matching scale
$\Lambda_{\rm match}$ at which an effective three-body description becomes
appropriate.  The second is the infrared dynamical scale
$\Lambda_{\rm 3B}$ generated inside the composite sector.

Below $\Lambda_{\rm match}$ the effective theory may contain genuine
three-body operators and the dimensionless holonomy phase $\theta_3$.  A
schematic local six-fermion operator would have the form
\begin{equation}
 {\cal L}_{3}
 \sim
 \frac{C_3}{\Lambda_{\rm match}^{5}}
 (\bar\psi\psi)^3,
\end{equation}
which only illustrates the role of the matching scale; it is not the
microscopic realization used later.

The ultimate origin of $\Lambda_{\rm match}$ is left unspecified.  If desired,
it may itself arise from ultraviolet dimensional transmutation.  For an
asymptotically free coupling $g_U$,
\begin{equation}
 \mu\frac{dg_U}{d\mu}
 =-\frac{b_U}{16\pi^2}g_U^3+\cdots,
\end{equation}
one can generate
\begin{equation}
 \Lambda_{\rm match}
 \sim
 M_{\rm Pl}
 \exp\left[-\frac{8\pi^2}{b_Ug_U^2(M_{\rm Pl})}\right].
 \label{eq:Lambda_match}
\end{equation}
This ultraviolet step is optional and is not derived from the three-body Dirac
model.  The dynamical three-body scale used in the low-energy theory will
instead be generated explicitly by the hidden strong dynamics introduced in
Sec.~\ref{sec:composite_model}.

\section{From configuration-space holonomy to internal holonomy}

In $(1+1)$ dimensions, removal of the center-of-mass coordinate leaves a
two-dimensional relative configuration space, and the triple-coincidence
point can support a nontrivial winding phase.
In $(3+1)$ dimensions the same configuration-space topology does not survive
unchanged, so the low-dimensional winding should not be transplanted
literally.

Instead, the $(1+1)$-dimensional result is used as a proof of concept which
a genuine three-body cyclic structure can carry nontrivial holonomy.
We then conjecture that in $(3+1)$ dimensions the corresponding phase
information can survive as an effective internal holonomy,
\begin{equation}
 W_3={\cal P}\exp\left(i\oint {\cal A}_3\right),
\end{equation}
whose Abelian reduction reproduces Eq.~(\ref{eq:W3}).

The internal space is not assumed to be an additional fundamental
spacetime dimension.  We define it minimally as the finite-dimensional Hilbert
space that retains the cyclic sector information of the three-body problem,
\begin{equation}
 {\cal H}_{\rm int}=\mathrm{span}\{|1\rangle,|2\rangle,|3\rangle\},
 \qquad
 \Psi(x)\in {\cal H}_{3+1}\otimes{\cal H}_{\rm int}.
 \label{eq:Hint}
\end{equation}
The labels $|i\rangle$ therefore denote internal three-body sectors, not
ordinary spatial positions.  Equivalently, the same construction may be
regarded as a three-point deconstruction of an auxiliary internal direction,
but no physical extra dimension is required by the model.

The number three is selected here by two related conditions.  First, the
microscopic motivation is a genuine three-body operator, not a generic
$N$-body interaction.  Second, three sites form the smallest oriented cycle
that supports an elementary loop variable built from three independent links,
\begin{equation}
 1\longrightarrow2\longrightarrow3\longrightarrow1.
\end{equation}
A four-site cycle is mathematically possible, but it introduces an additional
internal degree of freedom not required to represent the primitive three-body
holonomy.  We therefore adopt the body-number matching condition
\begin{equation}
 N_{\rm site}=N_{\rm body}
 \label{eq:body_site_matching}
\end{equation}
for the elementary holonomy sector.  In this restricted sense, three is not
chosen because a four-site theory is inconsistent; it is chosen because the
three-site cycle is the minimal faithful internal realization of the original
genuine three-body structure.

The conceptual order is therefore
\begin{align}
 \text{genuine three-body structure}
 &\rightarrow
 \text{three cyclic sectors} \nonumber \\
 &\rightarrow
 \text{three-site internal space} \nonumber \\
 &\rightarrow
 \text{holonomy}.
\end{align}

\section{Minimal three-site realization of the internal holonomy}
\label{sec:three_site}

The previous section motivates an internal holonomy in $(3+1)$ dimensions,
but does not by itself specify a microscopic origin for the corresponding
connection.  We therefore introduce a minimal deconstructed realization whose
purpose is narrower. The goal is to show explicitly that a threefold cyclic internal space
supports a gauge-invariant loop phase and that the same phase naturally
produces the leading $\cos\theta_3$ dependence used in the Higgs portal.

Consider three internal sites $i=1,2,3$ connected cyclically by link variables
 \begin{equation}
  U_{12},\qquad U_{23},\qquad U_{31}.
 \end{equation}
Under local rephasings
 \begin{equation}
  \psi_i\rightarrow e^{i\alpha_i}\psi_i,
 \end{equation}
the links transform as
 \begin{equation}
  U_{ij}\rightarrow e^{i(\alpha_i-\alpha_j)}U_{ij}.
 \end{equation}
The individual link phases are therefore basis dependent, whereas the product
around the closed three-site cycle is invariant,
\begin{equation}
 W_3\equiv U_{12}U_{23}U_{31}
 \rightarrow W_3.
 \label{eq:three_site_W}
\end{equation}
For an Abelian realization we may write
\begin{equation}
 W_3=e^{i\theta_3},\qquad
 \theta_3=\phi_{12}+\phi_{23}+\phi_{31}\pmod{2\pi}.
 \label{eq:theta_links}
\end{equation}
Thus the physical phase is associated with the closed three-site loop rather
than with any individual link.  This provides an explicit effective
realization of the internal holonomy motivated by the $(1+1)$-dimensional
three-body problem. 

It is useful to distinguish minimality from energetic stability.  The present
argument does not assert that an $N=3$ configuration has already been shown to
have a lower energy than every $N\ge4$ configuration.  Rather,
\begin{align}
 N=3
 =\min\{N\,|&\,\text{an elementary loop faithfully} \nonumber \\
 &\text{ represents the genuine three-body sector}\}.
 \label{eq:minimal_three}
\end{align}
An energetic preference for the three-site realization may emerge once the
microscopic cost of additional composite modes is specified, but such a
statement is model dependent. 

The same construction reproduces the threefold spectral form.  To avoid confusion
with the Standard-Model Higgs doublet $H$ introduced below, we denote the
operator acting on the three-dimensional cyclic internal space by
${\cal M}_{\rm cyc}$. It is an effective
Hermitian mass (or hopping) operator on ${\cal H}_{\rm int}$.  A minimal cyclic
hopping operator can be written as
\begin{equation}
 {\cal M}_{\rm cyc}
 =M_0\mathbf 1+
 \Delta\left(
 U_{12}|1\rangle\langle2|+
 U_{23}|2\rangle\langle3|+
 U_{31}|3\rangle\langle1|+{\rm h.c.}
 \right).
 \label{eq:Mcyc}
\end{equation}
Using the phase freedom of the three sites, the total holonomy may be
distributed uniformly,
\begin{equation}
 U_{12}=U_{23}=U_{31}=e^{i\theta_3/3}.
\end{equation}
In the site basis this gives
\begin{equation}
 {\cal M}_{\rm cyc}=M_0\mathbf 1+\Delta
 \begin{pmatrix}
 0&e^{i\theta_3/3}&e^{-i\theta_3/3}\\
 e^{-i\theta_3/3}&0&e^{i\theta_3/3}\\
 e^{i\theta_3/3}&e^{-i\theta_3/3}&0
 \end{pmatrix}.
 \label{eq:Mcyc_matrix}
\end{equation}
The $C_3$ eigenvectors
\begin{equation}
 |k\rangle=\frac{1}{\sqrt3}
 \begin{pmatrix}
 1\\ e^{2\pi ik/3}\\ e^{4\pi ik/3}
 \end{pmatrix},\qquad k=0,1,2,
\end{equation}
diagonalize Eq.~(\ref{eq:Mcyc_matrix}), with eigenvalues
\begin{equation}
 M^{[k]}=M_0+2\Delta
 \cos\left(\frac{\theta_3+2\pi k}{3}\right),
 \qquad k=0,1,2.
 \label{eq:three_site_spectrum}
\end{equation}
This is the same holonomy-dependent threefold form obtained in the effective
cyclic description of the $(1+1)$-dimensional model.  The agreement does not
identify the two constructions microscopically, but shows that the spectral
structure has a simple internal-loop realization.

The three-site loop also constrains the leading portal dependence.  A
Hermitian CP-even low-energy interaction may contain the gauge-invariant
combination
\begin{equation}
 W_3+W_3^\dagger
 =e^{i\theta_3}+e^{-i\theta_3}
 =2\cos\theta_3.
 \label{eq:WplusW}
\end{equation}
Hence the minimal portal operator can be written as
\begin{equation}
 V_{\rm portal}
 =\frac12\left[
 \kappa_0+\frac{\kappa_1}{2}
 (W_3+W_3^\dagger)
 \right]S_3^2H^\dagger H,
\end{equation}
which immediately becomes
\begin{equation}
 V_{\rm portal}
 =\frac12[\kappa_0+\kappa_1\cos\theta_3]
 S_3^2H^\dagger H.
 \label{eq:portal_from_loop}
\end{equation}
Higher windings generate
$(W_3^n+W_3^{\dagger n})=2\cos(n\theta_3)$, reproducing the general CP-even
Fourier expansion.  The previously phenomenological $\cos\theta_3$ portal is
therefore the leading invariant of the minimal three-site loop.

The resulting logical chain is
\begin{align}
 C_3\ {\rm cyclicity}
 &\longrightarrow
 W_3=U_{12}U_{23}U_{31} \nonumber \\
 &\longrightarrow
 W_3+W_3^\dagger=2\cos\theta_3 \nonumber \\
 &\longrightarrow
 \kappa_{H3}(\theta_3).
 \label{eq:three_site_chain}
\end{align}
This section should be read as an explicit $(3+1)$-dimensional effective
realization of the assumed internal phase.

\subsection{Why three? Minimality and repeated threefold structure}
\label{sec:why_three}

The role of the number three is central to the construction.  It is important,
however, to separate a minimality statement from a claim of absolute dynamical
uniqueness.  We do not claim that four- or higher-site theories are
inconsistent.  Rather, the hierarchy of elementary cyclic structures is
schematically
\begin{equation}
 \begin{array}{rcl}
 N=2 &:& \text{pairwise link structure,} \\
     &\quad &  \text{but no primitive three-body cycle},\\
 N=3 &:& \text{minimal faithful cycle } \\
     &\quad& \text{for a genuine three-body source},\\
 N>3 &:& \text{possible nonminimal higher-site extensions}.
 \end{array}
 \label{eq:why_three_hierarchy}
\end{equation}
Thus
\begin{align}
 3=\min\{N\,|\,N &\text{ supports the primitive cyclic} \nonumber \\
  &\text{realization of the genuine} \nonumber \\
  &\text{three-body holonomy} \}.
 \label{eq:three_minimality_statement}
\end{align}
This is the precise sense in which the phrase ``nature chooses three'' is used
in this paper.  It is a hypothesis of structural economy. Once a genuine
many-body holonomy is required, the smallest faithful internal realization is
chosen before additional degrees of freedom are introduced.

The significance of this minimal threefold structure is strengthened by its
reappearance at several logically distinct levels:
\begin{align}
 3_{\rm body}
 =3_{\rm site}
 &=3_{\rm flavor}
 =3_{\rm link}  \nonumber \\
 &\text{within the reference realization.}
 \label{eq:repeated_three}
\end{align}
The equal signs in Eq.~(\ref{eq:repeated_three}) denote matching of the
minimal threefold organization, not a proof that all four quantities have an
identical microscopic origin.  The sequence is instead
\begin{align}
 \text{three-body origin}
 &\rightarrow C_3 
 \rightarrow \text{three internal states} \nonumber \\
 &\rightarrow N_f=3
 \rightarrow \det\Sigma\text{ cubic}.
 \label{eq:three_sequence}
\end{align}
For $N_f=4$, by contrast, the determinant is quartic and naturally belongs to
a four-link/higher-body extension.  The three-flavor theory is special because
its standard determinant invariant has exactly the order of the primitive
three-link cycle.  This repeated matching is the main reason for emphasizing
three rather than treating it as an incidental numerical choice.

\section{Composite three-body sector as the reference model}
\label{sec:composite_model}

We now give an explicit $(3+1)$-dimensional realization of the three-site
sector.  The purpose is to
show that the internal holonomy and the associated physical scale can arise
from ordinary QCD-like strong dynamics without introducing an elementary
three-body scalar by hand.

\subsection{Hidden $SU(3)_H$ dynamics with three singlet flavors}
\label{sec:hidden_example}

Consider a hidden confining gauge group
\begin{equation}
 G_H=SU(3)_H
\end{equation}
with three Dirac fermion flavors
\begin{equation}
 \chi_i^a,\qquad i=1,2,3,\qquad a=1,2,3.
\end{equation}
All $\chi_i$ are singlets under the Standard Model gauge group,
\begin{equation}
 \chi_i:\quad (\mathbf 1,\mathbf 1)_0
 \quad\text{under }SU(3)_C\times SU(2)_L\times U(1)_Y.
 \label{eq:chi_SM_singlet}
\end{equation}
The hidden-color index $a$ and the flavor/site index $i$ have distinct roles.
The latter is fixed first by the threefold holonomy structure, whereas the
former supplies a QCD-like confining reference dynamics.  Thus the hidden
$SU(3)_H$ theory should not be read as the reason that three was chosen.
Rather, once the primitive three-body sector has selected a three-state
organization, $SU(3)_H$ with $N_f=3$ provides an economical composite
realization in which asymptotic freedom, three-body confinement, and a cubic
determinant coexist.  The equality of the hidden-color and flavor dimensions
is therefore a structurally suggestive matching, not a derivation of one from
the other.

At one loop, the running is asymptotically free for the parameter choice used below
\cite{GrossWilczek1973,Politzer1973},
\begin{equation}
 \mu\frac{dg_H}{d\mu}
 =-\frac{b_H}{16\pi^2}g_H^3+\cdots,
 \qquad
 b_H=\frac{11}{3}N_H-\frac{2}{3}N_f.
 \label{eq:hidden_beta}
\end{equation}
For $N_H=N_f=3$,
\begin{equation}
 b_H=9>0,
 \label{eq:bh9}
\end{equation}
so the theory is asymptotically free and generates
\begin{equation}
 \Lambda_H
 =\Lambda_{\rm match}
 \exp\left[-\frac{8\pi^2}
 {9g_H^2(\Lambda_{\rm match})}\right].
 \label{eq:LambdaH}
\end{equation}
The scale $\Lambda_H$ will be identified, up to nonperturbative order-one
factors, with the dynamical three-body scale.

\subsection{Composite matrix, determinant, and the cyclic invariant}

Strong dynamics produces meson-like operators
\begin{equation}
 \Sigma_{ij}\sim\bar\chi_{Rj}\chi_{Li}.
 \label{eq:Sigma_composite}
\end{equation}
The field $\Sigma$ should be understood as a low-energy composite
order parameter of the confining hidden sector, analogous to the
mesonic matrix used in QCD-like effective descriptions.  Hidden
composite sectors communicating with Higgs dynamics have also been
studied in related contexts \cite{NelsonParkWalker2019}.
In the chiral limit the matrix transforms schematically as
$\Sigma\rightarrow U_L\Sigma U_R^\dagger$.  A standard three-flavor
low-energy potential may contain an anomalous determinant interaction
\cite{tHooft1976,PisarskiWilczek1984},
\begin{equation}
 V_A
 =-\left[c_Ae^{i\delta_A}\det\Sigma+\mathrm{h.c.}\right],
 \label{eq:det_potential}
\end{equation}
where the canonically normalized field has mass dimension one and therefore
$c_A$ has mass dimension one.  The phase $\delta_A$ denotes a possible
intrinsic phase of the anomalous determinant coupling in the hidden strong
sector.  It is conceptually distinct from the inherited three-body holonomy
$\theta_3$: the former belongs to the coefficient of the local determinant
operator, whereas the latter specifies the cyclic orientation of the
three-state sector.  Only the rephasing-invariant combination
$\theta_3+\delta_A$ enters the determinant energy below.  In a CP-conserving
reference limit one may set $\delta_A=0$, but we retain it in the formulas to
make the phase structure explicit.

For a $3\times3$ matrix,
\begin{align}
 \det\Sigma={}&
 \Sigma_{11}\Sigma_{22}\Sigma_{33}
 +\Sigma_{12}\Sigma_{23}\Sigma_{31}
 +\Sigma_{13}\Sigma_{21}\Sigma_{32}
 \nonumber\\
 &-\Sigma_{13}\Sigma_{22}\Sigma_{31}
 -\Sigma_{12}\Sigma_{21}\Sigma_{33}
 -\Sigma_{11}\Sigma_{23}\Sigma_{32}.
 \label{eq:det_expansion}
\end{align}
Hence the oriented three-link product appears automatically,
\begin{equation}
 \det\Sigma\supset
 \Sigma_{12}\Sigma_{23}\Sigma_{31}.
 \label{eq:det_contains_cycle}
\end{equation}
Defining
\begin{equation}
 \Phi_{12}=\Sigma_{12},\qquad
 \Phi_{23}=\Sigma_{23},\qquad
 \Phi_{31}=\Sigma_{31},
\end{equation}
and writing $\Phi_{ij}=\rho_{ij}e^{i\phi_{ij}}$, their cyclic product is
\begin{align}
 {\cal C}_3
 =\Phi_{12}\Phi_{23}\Phi_{31}
 =|{\cal C}_3|e^{i\theta_3},
 \qquad
 \theta_3&=\phi_{12}+\phi_{23}+\phi_{31} \nonumber \\
 & \pmod{2\pi}  .
 \label{eq:composite_holonomy}
\end{align}
Under site rephasings the individual phases shift, but their sum does not.
Thus the determinant term provides a conventional composite-theory origin for
the same rephasing-invariant cyclic phase introduced in
Sec.~\ref{sec:three_site}.

The special role of three is now visible in a second way.  For $N_f$ flavors,
$\det\Sigma$ is an $N_f$th-order invariant.  Thus $N_f=2$ gives a quadratic
determinant, $N_f=3$ gives the cubic invariant in
Eq.~(\ref{eq:det_contains_cycle}), and $N_f=4$ gives a quartic invariant.  The
three-flavor theory is therefore the case in which the determinant interaction
has exactly the order required by the primitive three-link cycle.

\subsection{Cyclic alignment and vacuum stability}
\label{sec:cyclic_vacuum}

The determinant interaction contains the desired cyclic product, but by itself
it does not force a flavor-symmetric condensate to point in the off-diagonal
cyclic direction.  Before discussing that orientation, it is useful to state
why a nonzero condensate is expected at all.  The hidden theory is
asymptotically free and becomes strongly coupled at $\Lambda_H$.  By analogy
with QCD-like dynamics, the attractive strong interaction is assumed to
trigger spontaneous chiral symmetry breaking and to generate a bilinear
order parameter
\begin{equation}
 \langle\bar\chi_{Rj}\chi_{Li}\rangle\neq0,
 \qquad
 \langle\Sigma_{ij}\rangle\neq0,
 \label{eq:Sigma_condensate_origin}
\end{equation}
with characteristic magnitude set by the confinement scale.  In the
low-energy linear-sigma description this dynamical assumption is represented
by parameters for which the effective potential develops a minimum at
$\Sigma\neq0$. The simplest reference choice is $m_\Sigma^2<0$.  The precise
nonperturbative relation between $v_\Sigma$ and $\Lambda_H$ is not calculated
here and is absorbed later into an order-one coefficient.  Thus the strong
dynamics supplies the magnitude of the condensate, while the holonomy
input is used only to specify its cyclic orientation.

We therefore keep the mapping from the original three-body holonomy as an
explicit spurion that selects this internal orientation.  Define
\begin{equation}
 C_3(\theta_3)=
 \begin{pmatrix}
 0&e^{i\phi_{12}}&0\\
 0&0&e^{i\phi_{23}}\\
 e^{i\phi_{31}}&0&0
 \end{pmatrix},
 \qquad
 \det C_3(\theta_3)=e^{i\theta_3}.
 \label{eq:C3theta}
\end{equation}
A minimal reference potential is
\begin{align}
 V_\Sigma={}&
 m_\Sigma^2\Tr(\Sigma^\dagger\Sigma)
 +\lambda_1[\Tr(\Sigma^\dagger\Sigma)]^2
 +\lambda_2\Tr[(\Sigma^\dagger\Sigma)^2]
 \nonumber\\
 &-\left[c_Ae^{i\delta_A}\det\Sigma+\mathrm{h.c.}\right]
 \nonumber\\
 &+\mu_A^2\left[
 \Tr(\Sigma^\dagger\Sigma)
 -\frac13
 \left|\Tr(C_3^\dagger\Sigma)\right|^2
 \right].
 \label{eq:VSigma_full}
\end{align}
The last line is non-negative by the Cauchy--Schwarz inequality and vanishes
for $\Sigma\propto C_3$.  It should be interpreted as a spurion-induced
alignment term. The strong dynamics determines the magnitude of the order
parameter, while the inherited three-body holonomy specifies its cyclic
orientation.

Consider the cyclic vacuum ansatz
\begin{equation}
 \langle\Sigma\rangle
 =v_\Sigma C_3(\theta_3).
 \label{eq:cyclic_vev}
\end{equation}
Then
\begin{align}
 \Tr(\Sigma^\dagger\Sigma)=3v_\Sigma^2,
 &\qquad
 \Tr[(\Sigma^\dagger\Sigma)^2]=3v_\Sigma^4, \nonumber \\
 &\qquad
 \det\Sigma=v_\Sigma^3e^{i\theta_3},
\end{align}
and the alignment term vanishes.  Defining
\begin{equation}
 B\equiv3\lambda_1+\lambda_2,
 \qquad
 c_\theta\equiv|c_A|\cos(\theta_3+\delta_A),
\end{equation}
the radial potential becomes
\begin{equation}
 V(v_\Sigma)
 =3m_\Sigma^2v_\Sigma^2
 +3Bv_\Sigma^4
 -2c_\theta v_\Sigma^3.
 \label{eq:Vrad_vSigma}
\end{equation}
The nonzero stationary points satisfy
\begin{equation}
 2Bv_\Sigma^2-c_\theta v_\Sigma+m_\Sigma^2=0,
\end{equation}
so
\begin{equation}
 v_{\Sigma,\pm}
 =\frac{c_\theta\pm
 \sqrt{c_\theta^2-8Bm_\Sigma^2}}{4B}.
 \label{eq:vSigma_pm}
\end{equation}
For $B>0$ and $m_\Sigma^2<0$, the "$+$" branch is always real and positive and
obeys
\begin{equation}
 \left.\frac{d^2V}{dv_\Sigma^2}\right|_{v_{\Sigma,+}}
 =6v_{\Sigma,+}
 \sqrt{c_\theta^2-8Bm_\Sigma^2}>0,
\end{equation}
so it is a radial minimum.  Even for $m_\Sigma^2>0$, the cubic determinant
term can generate a nonzero local minimum when the discriminant is positive;
the nonzero vacuum lies below the origin when
\begin{equation}
 m_\Sigma^2<\frac{c_\theta^2}{9B}
 \qquad(c_\theta>0).
 \label{eq:first_order_condition}
\end{equation}
We do not rely on this latter possibility in the reference model.

Angular stability follows directly from the alignment term.  Writing
$\Sigma=v_\Sigma C_3e^{iX}$ with $X=X^\dagger$ and $\Tr X=0$,
\begin{equation}
 V_{\rm align}
 =\mu_A^2v_\Sigma^2\Tr X^2+O(X^3),
\end{equation}
so $\mu_A^2>0$ stabilizes fluctuations away from the cyclic orientation.

If $\theta_3$ itself were treated as a dynamical angular field, the determinant
term would minimize at $\theta_3=-\delta_A$ modulo $2\pi$.  In the present
construction, however, $\theta_3$ is treated as a fixed holonomy-sector
parameter inherited from the ultraviolet three-body structure.  This
distinction is essential if a nontrivial fixed value of $\theta_3$ is to be
retained in the low-energy portal.

\subsection{Radial composite mode and three-body scale}

For the cyclic vacuum of Eq.~(\ref{eq:cyclic_vev}), the ordinary flavor-singlet
projection $\Re\Tr\Sigma$ vanishes because $\Tr C_3=0$.  The radial mode
must therefore be defined along the cyclic direction itself.  We write
\begin{equation}
 \Sigma(x)
 =\frac{S_3(x)}{\sqrt6}
 C_3(\theta_3)+\cdots,
 \label{eq:S3_definition}
\end{equation}
or equivalently
\begin{equation}
 S_3
 =\sqrt{\frac23}
 \Re\Tr[C_3^\dagger(\theta_3)\Sigma].
 \label{eq:S3_projection}
\end{equation}
Then
\begin{equation}
 \Tr(\Sigma^\dagger\Sigma)=\frac12S_3^2,
 \qquad
 \langle S_3\rangle=\sqrt6\,v_\Sigma,
\end{equation}
and the quartic term along this direction is
\begin{equation}
 \frac{\lambda_3}{4}S_3^4,
 \qquad
 \lambda_3\equiv\frac{B}{3}
 =\frac{3\lambda_1+\lambda_2}{3}.
 \label{eq:lambda3_radial}
\end{equation}
Moreover,
\begin{equation}
 \det\Sigma
 =\frac{S_3^3}{6\sqrt6}e^{i\theta_3}.
\end{equation}
Consequently the determinant contribution along the cyclic direction is
\begin{equation}
 V_{\rm hol}(S_3,\theta_3)
 =-\frac{|c_A|}{3\sqrt6}
 S_3^3\cos(\theta_3+\delta_A).
 \label{eq:Vhol_S3}
\end{equation}
At the vacuum we define
\begin{equation}
 \Lambda_{\rm 3B}^4
 \equiv
 2|c_A|v_\Sigma^3
 =\frac{|c_A|}{3\sqrt6}
 \langle S_3\rangle^3.
 \label{eq:Lambda3B}
\end{equation}
If the hidden theory is governed by a single strong scale,
$v_\Sigma\sim\Lambda_H$ and $|c_A|\sim\Lambda_H$, giving parametrically
\begin{equation}
 \Lambda_{\rm 3B}\sim\Lambda_H.
 \label{eq:Lambda3B_LambdaH}
\end{equation}
The proportionality coefficient is nonperturbative and is not predicted in
this work.

We parameterize the radial condensate by
\begin{equation}
 \langle S_3\rangle=c_S\Lambda_H,
 \qquad c_S=O(1).
 \label{eq:S3_condensate}
\end{equation}
This definition absorbs the nonperturbative normalization of the composite
field into $c_S$.

\subsection{Higgs potential and the holonomy-dependent portal}

Before writing the portal interaction, it is useful to distinguish the two
scalar quantities that enter the low-energy theory.  The field
$H$ is the ordinary Standard-Model electroweak Higgs doublet, transforming as
$({\bf 1},{\bf 2})_{1/2}$ under
$SU(3)_C\times SU(2)_L\times U(1)_Y$.  By contrast, $S_3$ is a real
Standard-Model-singlet composite radial mode of the hidden strong sector.  It
is not an elementary Higgs-like field: it is defined by projecting the
composite matrix $\Sigma$ onto the cyclic condensate direction,
Eq.~(\ref{eq:S3_definition}), and its expectation value
$\langle S_3\rangle=\sqrt6\,v_\Sigma$ is generated by the hidden
strong dynamics.  Thus the roles of the two fields are distinct.
$S_3$ carries the dynamically generated three-body scale and its holonomy
dependence, whereas $H$ is the visible-sector Higgs field whose quadratic
coefficient is shifted through the portal coupling.

Schematically, the mechanism is
\begin{align}
 \text{hidden strong dynamics}
 &\;\longrightarrow\;\langle S_3\rangle\neq0 \nonumber \\
 &\;\longrightarrow\;
 m_{H,{\rm eff}}^2(\theta_3) \nonumber \\
 &\;\longrightarrow\;
 \langle H\rangle\neq0,
 \label{eq:S3_Higgs_chain}
\end{align}
when the induced Higgs quadratic coefficient becomes negative.  The portal
therefore transmits the dynamically generated hidden-sector scale to the
electroweak Higgs sector.

To make the electroweak-breaking mechanism explicit, we first write the Higgs
potential before coupling it to the composite sector.  In the reference
construction we take
\begin{equation}
 V_H(H)=m_{H,0}^2 H^\dagger H
 +\lambda_H(H^\dagger H)^2,
 \qquad \lambda_H>0,
 \label{eq:bare_Higgs_potential}
\end{equation}
where $m_{H,0}^2$ denotes the Higgs quadratic parameter in the low-energy
matching theory.  The purpose of the composite sector is to shift this
quadratic coefficient dynamically.  The simplest portal before radial
projection is
\begin{equation}
 V_{\rm portal}
 =\kappa_{H3}(\theta_3)
 H^\dagger H\,\Tr(\Sigma^\dagger\Sigma),
\end{equation}
which becomes
\begin{equation}
 V_{\rm portal}
 =\frac12\kappa_{H3}(\theta_3)S_3^2H^\dagger H.
 \label{eq:portal_radial}
\end{equation}
Thus the Higgs-dependent part of the scalar potential is
\begin{equation}
 V_H+V_{\rm portal}
 =\left[m_{H,0}^2+
 \frac12\kappa_{H3}(\theta_3)S_3^2\right]H^\dagger H
 +\lambda_H(H^\dagger H)^2.
 \label{eq:Higgs_plus_portal}
\end{equation}
For a CP-even low-energy observable, the leading periodic loop invariant is
$W_3+W_3^\dagger=2\cos\theta_3$, so
\begin{equation}
 \kappa_{H3}(\theta_3)
 =\kappa_0+\kappa_1\cos\theta_3.
 \label{eq:kappa_theta}
\end{equation}
The quartic potential is bounded from below provided
\begin{equation}
 \lambda_H>0,\qquad
 \lambda_3>0,\qquad
 \kappa_{H3}(\theta_3)>-2\sqrt{\lambda_H\lambda_3},
 \label{eq:vacuum_stability}
\end{equation}
where $\lambda_3$ denotes the radial quartic after projection onto the
$C_3$ direction.

Once $S_3$ condenses, Eq.~(\ref{eq:Higgs_plus_portal}) becomes
\begin{equation}
 V_H^{\rm eff}(H)
 =m_{H,\rm eff}^2(\theta_3)H^\dagger H
 +\lambda_H(H^\dagger H)^2,
 \label{eq:Higgs_effective_potential}
\end{equation}
with
\begin{equation}
 m_{H,\rm eff}^2(\theta_3)
 =m_{H,0}^2
 +\frac12\kappa_{H3}(\theta_3)
 \langle S_3\rangle^2.
 \label{eq:mH_from_S3}
\end{equation}
Electroweak symmetry breaking occurs when
$m_{H,\rm eff}^2(\theta_3)<0$.  With the convention
$\langle H\rangle=(0,v_{\rm EW}/\sqrt2)^T$, minimization gives
\begin{equation}
 v_{\rm EW}^2
 =-\frac{m_{H,\rm eff}^2(\theta_3)}{\lambda_H}
 =-\frac{1}{\lambda_H}
 \left[
 m_{H,0}^2+
 \frac12\kappa_{H3}(\theta_3)\langle S_3\rangle^2
 \right].
 \label{eq:v_general_from_S3}
\end{equation}
In the classically scale-invariant reference limit $m_{H,0}^2=0$, this
reduces to
\begin{equation}
 v_{\rm EW}
 =\langle S_3\rangle
 \sqrt{-\frac{\kappa_{H3}(\theta_3)}{2\lambda_H}},
 \qquad \kappa_{H3}(\theta_3)<0,
 \label{eq:v_from_S3}
\end{equation}
which is the form used below.  Writing the more general expression first makes
clear that the portal does not constitute the whole Higgs potential. Rather,
it dynamically shifts the coefficient of the ordinary Higgs quadratic term.
Specializing henceforth to the reference limit $m_{H,0}^2=0$, combining
Eqs.~(\ref{eq:LambdaH}), (\ref{eq:S3_condensate}), and
(\ref{eq:kappa_theta}), we obtain the central relation
\begin{equation}
 v_{\rm EW}
 =c_S\Lambda_{\rm match}
 \exp\left[-\frac{8\pi^2}
 {9g_H^2(\Lambda_{\rm match})}\right]
 \sqrt{-\frac{\kappa_0+\kappa_1\cos\theta_3}
 {2\lambda_H}}.
 \label{eq:master_composite}
\end{equation}
Thus the hidden strong dynamics generates the physical scale, whereas the
holonomy controls its transmission into the Higgs sector.

\subsection{Interpretation of the reference realization}

The logical roles are
\begin{equation}
 \begin{array}{ccl}
 \text{three-body origin}
 &\longrightarrow& C_3(\theta_3)\ \text{orientation},\\[1mm]
 SU(3)_H\ \text{running}
 &\longrightarrow& \Lambda_H\sim\Lambda_{\rm 3B},\\[1mm]
 N_f=3
 &\longrightarrow& \det\Sigma\ \text{cubic},\\[1mm]
 \det\Sigma
 &\longrightarrow& \Phi_{12}\Phi_{23}\Phi_{31},\\[1mm]
 \theta_3
 &\longrightarrow& \kappa_{H3}(\theta_3),\\[1mm]
 \kappa_{H3}<0
 &\longrightarrow& \text{electroweak symmetry breaking}.
 \end{array}
 \label{eq:roles_hidden}
\end{equation}
The hidden baryon operator
\begin{equation}
 B_H\sim\epsilon_{abc}\chi^a\chi^b\chi^c
\end{equation}
may be stable if an appropriate hidden baryon number is exact or sufficiently
long lived.  It could then provide a composite dark-matter candidate, but this
possibility is not used in the hierarchy mechanism and is left for separate
study.

\section{Dynamical electroweak-scale selection and parameter sensitivity}
\label{sec:hierarchy}

The scale structure of the explicit reference model is
\begin{equation}
 M_{\rm Pl}
 >
 \Lambda_{\rm match}
 >
 \Lambda_H\sim\Lambda_{\rm 3B}
 >
 M_{\rm EW}.
 \label{eq:full_scale_chain}
\end{equation}
The first step is ultraviolet and remains unspecified.  The second is generated
by hidden strong running, and the final step is controlled by the
holonomy-dependent portal.  In the reference EFT, no elementary scalar mass
of order $\Lambda_{\rm match}$ is introduced with a direct tree-level
coupling to the Higgs.  Possible threshold corrections generated by heavy
states in an ultraviolet completion are not determined by the present toy
construction.

If $\Lambda_{\rm match}$ itself is generated from the Planck scale by an
ultraviolet exponential,
\begin{equation}
 \Lambda_{\rm match}=M_{\rm Pl}e^{-X_{\rm UV}},
\end{equation}
and
\begin{equation}
 X_H
 =\frac{8\pi^2}{9g_H^2(\Lambda_{\rm match})},
\end{equation}
then Eq.~(\ref{eq:master_composite}) becomes
\begin{equation}
 M_{\rm EW}
 \sim
 c_S M_{\rm Pl}
 e^{-X_{\rm UV}}e^{-X_H}
 \sqrt{-\frac{\kappa_{H3}(\theta_3)}{2\lambda_H}}.
 \label{eq:full_hierarchy_composite}
\end{equation}
The logarithmic hierarchy is therefore approximately
\begin{equation}
 \ln\frac{M_{\rm Pl}}{M_{\rm EW}}
 \simeq
 X_{\rm UV}+X_H-\ln c_S
 -\frac12\ln\left|
 \frac{-\kappa_{H3}(\theta_3)}{2\lambda_H}
 \right|.
 \label{eq:log_hierarchy_composite}
\end{equation}
The hierarchy is not eliminated. It is reorganized into dimensional
transmutation and a dimensionless portal rather than an explicit microscopic
mass ratio of order $10^{-17}$.

If $\Lambda_{\rm match}\sim M_{\rm Pl}$, the hidden exponential alone would
need roughly
\begin{equation}
 X_H\sim\ln\frac{M_{\rm Pl}}{M_{\rm EW}}\sim38,
\end{equation}
up to order-one portal factors.  With $b_H=9$ this corresponds to
\begin{equation}
 g_H^2(\Lambda_{\rm match})
 \sim\frac{8\pi^2}{9\times38},
\qquad
 g_H(\Lambda_{\rm match})\sim0.48,
\end{equation}
showing that an exponentially large scale separation need not require an
extremely small ultraviolet coupling.

\begin{figure}[t]
 \centering
 \includegraphics[width=0.82\linewidth]{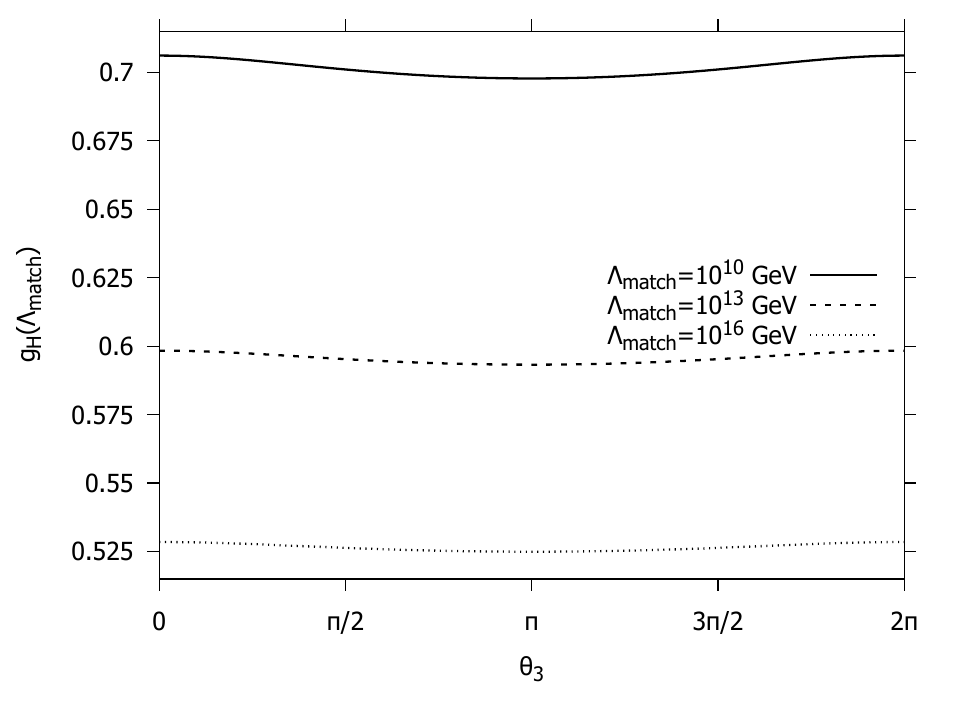}
 \caption{
 Required value of $g_H(\Lambda_{\rm match})$ as a function of the holonomy
 angle $\theta_3$ for several choices of $\Lambda_{\rm match}$.  The curves
 use Eq.~(\ref{eq:master_composite}) with $v_{\rm EW}=246\,\mathrm{GeV}$,
 $c_S=1$, $b_H=9$, $\lambda_H=0.129$, and
 $\kappa_{H3}(\theta_3)=-0.50+0.20\cos\theta_3$.  The required coupling is
 moderate over the displayed range, illustrating dimensional transmutation in
 the explicit $N_H=N_f=3$ reference model.
 }
 \label{Fig1-BW}
\end{figure}

A logarithmic sensitivity measure gives
\begin{equation}
 \Delta_{g_H}
 \equiv
 \left|
 \frac{\partial\ln M_{\rm EW}}
 {\partial\ln g_H}
 \right|
 =2X_H,
\end{equation}
whereas
\begin{equation}
 \Delta_{\kappa}
 =\left|
 \frac{\partial\ln M_{\rm EW}}
 {\partial\ln|\kappa_{H3}|}
 \right|
 =\frac12.
\end{equation}
The resulting dependence of the required hidden coupling on the holonomy angle
is shown in Fig.~\ref{Fig1-BW} for several representative values of
$\Lambda_{\rm match}$. 
As seen in Fig.~\ref{Fig1-BW}, the required $g_H(\Lambda_{\rm match})$ 
remains moderate over the full range of $\theta_3$, 
while its dependence on the holonomy angle is comparatively mild. 
Thus the model replaces an explicitly tiny dimensionful scale ratio 
by an exponentially generated hierarchy by a domensionless hidden coupling. 
In particular,
$\Delta_{g_H}=2X_H\simeq 76$ for $X_H\simeq38$, so the mechanism should not
be described as eliminating all parameter sensitivity.  Its more limited
naturalness statement is that the hierarchy is generated by dimensional
transmutation from moderate dimensionless inputs rather than by imposing a
fundamental mass ratio of order $10^{-17}$.

\subsection{Representative one-loop radiative check}
\label{sec:one_loop_check}

To test whether the tree-level scale-selection mechanism is immediately
destroyed by leading radiative effects, we consider a simplified radial
two-field potential,
\begin{equation}
 V_0(h,S_3)=
 \frac{\lambda_H}{4}h^4
 -\frac{\mu_3^2}{2}S_3^2
 +\frac{\lambda_3}{4}S_3^4
 +\frac{\kappa_{H3}}{4}h^2S_3^2,
 \label{eq:V0_oneloop}
\end{equation}
where no independent tree-level Higgs mass term is introduced.  The
one-loop Coleman--Weinberg contribution is written as
\cite{ColemanWeinberg1973,GildenerWeinberg1976}
\begin{equation}
 V_{\rm CW}^{(1)}
 =
 \sum_i
 \frac{n_i}{64\pi^2}m_i^4(h,S_3)
 \left[
 \ln\frac{m_i^2(h,S_3)}{Q^2}-c_i
 \right],
 \label{eq:VCW}
\end{equation}
with the dominant Standard-Model contributions from
$i=t,W,Z$ together with the two radial scalar eigenmodes.  We use
$c_i=3/2$ for scalars and fermions and $c_i=5/6$ for gauge bosons.

As a representative benchmark, take
\begin{equation}
 \langle S_3\rangle=1~{\rm TeV},\qquad
 \lambda_H=0.129,\qquad
 \lambda_3=0.20,
\end{equation}
and choose the portal coupling so that the tree-level minimum lies at
$v_{\rm EW}=246~{\rm GeV}$,
\begin{equation}
 \kappa_{H3}
 =
 -\frac{2\lambda_Hv_{\rm EW}^2}{\langle S_3\rangle^2}
 \simeq -1.56\times10^{-2}.
 \label{eq:kappa_benchmark}
\end{equation}
Using $Q=246~{\rm GeV}$, $g=0.653$, $g'=0.358$, and $y_t=0.995$, a direct
minimization of $V_0+V_{\rm CW}^{(1)}$ gives, for this benchmark,
\begin{equation}
 h_{\rm min}\simeq219~{\rm GeV},\qquad
 S_{3,{\rm min}}\simeq996~{\rm GeV}.
 \label{eq:oneloop_minimum}
\end{equation}
Thus the leading one-loop correction shifts the electroweak minimum by
roughly ten percent while leaving it in the electroweak-scale region. The
strong-sector condensate changes only at the sub-percent level.

This estimate is intentionally limited.  The residual dependence on the
renormalization scale $Q$ is not negligible in a fixed-coupling
Coleman--Weinberg calculation, and a quantitative statement of radiative
stability requires RG improvement of
$\lambda_H$, $\lambda_3$, $\kappa_{H3}$, the gauge couplings, and the top
Yukawa coupling.  Equation~(\ref{eq:oneloop_minimum}) should therefore be
read only as evidence that the selected electroweak minimum is not
immediately erased by the leading one-loop terms, not as a proof of complete
ultraviolet naturalness.

\begin{figure}[t]
 \centering
 \includegraphics[width=0.78\linewidth]{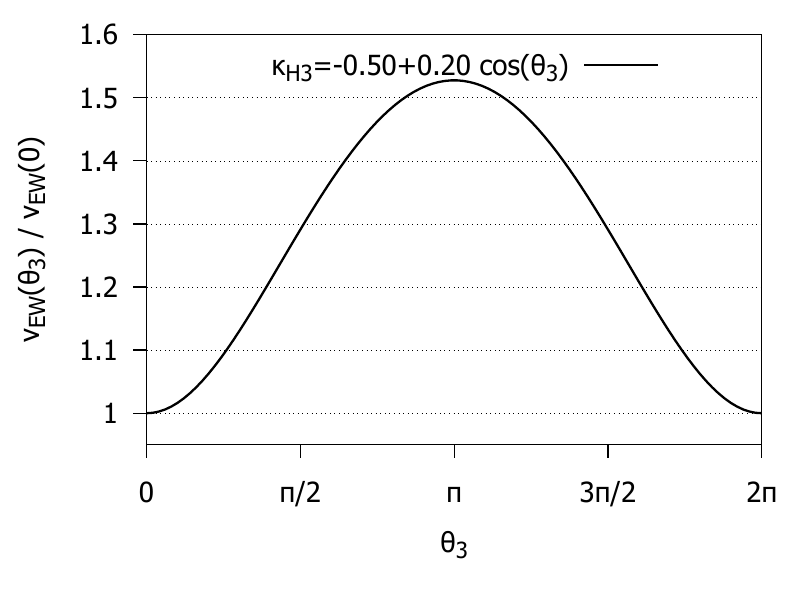}
 \caption{
 Holonomy modulation of the electroweak scale for
 $\kappa_{H3}(\theta_3)=-0.50+0.20\cos\theta_3$, with the hidden dynamical
 scale held fixed.  The figure isolates the role of the holonomy from the
 exponential strong-sector hierarchy.
 }
 \label{Fig2-BW}
\end{figure}

The direct role of the holonomy is clearer if the strong scale is held fixed.
From Eq.~(\ref{eq:v_from_S3}),
\begin{equation}
 \frac{v_{\rm EW}(\theta_3)}{v_{\rm EW}(0)}
 =\sqrt{
 \frac{-\kappa_0-\kappa_1\cos\theta_3}
 { -\kappa_0-\kappa_1}}
 \label{eq:v_ratio}
\end{equation}
for the parameter region in which the numerator and denominator correspond to
negative portal couplings. The corresponding holonomy modulation is displayed 
in Fig.~\ref{Fig2-BW}.
With $\kappa_0=-0.50$ and $\kappa_1=0.20$, the
ratio rises from unity at $\theta_3=0$ to approximately $1.53$ at
$\theta_3=\pi$.  The exponential hierarchy is therefore controlled mainly by
the strong running, while the holonomy provides an order-one modulation of the
final transmission to the Higgs sector.

\section{Optional trinification interpretation}
\label{sec:trinification}

The core scale-selection mechanism does not require trinification.  Once the
minimal threefold internal organization has emerged, however, it is natural
to ask whether it can be embedded in a continuous threefold gauge structure.
A possible phenomenological extension is
\begin{equation}
 G_{333}=SU(3)_C\times SU(3)_L\times SU(3)_R,
\end{equation}
as in trinification models \cite{BabuHePakvasa1986,Stech2014}.  Such an
embedding is not derived from the $(1+1)$-dimensional holonomy and should not
be regarded as an ingredient of the hierarchy mechanism.

Structurally, the three cyclic composite links may be associated
schematically with trinification bifundamentals,
\begin{equation}
 \Phi_{12}\leftrightarrow(3,\bar 3,1),\quad
 \Phi_{23}\leftrightarrow(1,3,\bar 3),\quad
 \Phi_{31}\leftrightarrow(\bar 3,1,3),
\end{equation}
whose cyclic product closes to a singlet.  This makes a subsequent
$SU(3)\rightarrow SU(2)\times U(1)$ reduction compatible with the same
threefold organization \cite{Yoshikawa2026SU3}.  The point is only structural.
Three is selected before the larger gauge symmetry is postulated.  A full
trinification scalar
sector, its symmetry-breaking chain, and its phenomenology are beyond the
scope of the present work.

\section{What is derived and what is assumed}

\paragraph{Established within the $(1+1)$-dimensional model.}
The operator $Q_3=\alpha_1\alpha_2\alpha_3$, the holonomy
$W_3=e^{i\theta_3Q_3}$, the opposite $Q_3=\pm1$ phases in the massless
limit, mass-induced sector mixing, and the threefold cyclic spectral structure
belong to the low-dimensional Dirac framework
\cite{Yoshikawa2026phase,Yoshikawa2026family}.  Within the same cyclic
three-state description, the special point $\theta_3=0$ produces a
$3\to2+1$ splitting whose traceless mass component is proportional to $T_8$;
its continuous commutant is therefore $SU(2)\times U(1)$
\cite{Yoshikawa2026SU3}.  This is a result of the low-dimensional toy
construction and should be distinguished from the four-dimensional composite
realization introduced below.

\paragraph{Assumed map to $(3+1)$ dimensions.}
The survival of the low-dimensional phase as a finite internal degree of
freedom is not derived from the original configuration-space topology.  The
spurion $C_3(\theta_3)$ is the explicit modeling assumption that carries the
three-body cyclic information into the four-dimensional composite theory.
The body-number matching condition $N_{\rm site}=N_{\rm body}$ is likewise a
minimality prescription for the elementary sector rather than a universal
law.

\paragraph{Derived inside the composite reference EFT.}
Once hidden $SU(3)_H$ with $N_f=3$ is chosen, asymptotic freedom gives
$b_H=9$ at one loop and dimensional transmutation gives
Eq.~(\ref{eq:LambdaH}).  The three-flavor determinant is necessarily cubic
and contains $\Sigma_{12}\Sigma_{23}\Sigma_{31}$.  Given the alignment
potential in Eq.~(\ref{eq:VSigma_full}), the cyclic vacuum
$\langle\Sigma\rangle=v_\Sigma C_3$ is an angular minimum for
$\mu_A^2>0$, while Eq.~(\ref{eq:vSigma_pm}) gives its radial stationary
points.  The definitions of $S_3$, $\Lambda_{\rm 3B}$, and the portal then
lead directly to Eq.~(\ref{eq:master_composite}).

\paragraph{Nonperturbative quantities not calculated here.}
The coefficients $m_\Sigma^2$, $\lambda_{1,2}$, $c_A$, $\mu_A^2$, and $c_S$
are not derived from first-principles hidden-QCD dynamics in this work.  In
particular, the sign and magnitude of the alignment term are treated as the
low-energy imprint of the ultraviolet three-body spurion.  A lattice or other
nonperturbative treatment would be required to determine the composite
spectrum and matching coefficients quantitatively.

\paragraph{Ultraviolet uncertainty.}
The microscopic degrees of freedom above $\Lambda_{\rm match}$ and the
origin of that matching scale are left unspecified.  Trinification, when
discussed, is an optional embedding rather than an ingredient of the core
composite mechanism.

\section{Discussion}

The revised construction separates two roles that were intertwined in the
more phenomenological version.  The three-body sector provides the cyclic
orientation, while strong dynamics provides the magnitude of the physical
scale. Three-body holonomy determines orientation,
 $\qquad $ strong dynamics generates scale.
The logical chain is
\begin{equation}
 \begin{array}{c}
 (1+1){\rm D\ genuine\ three\! -\! body\ holonomy}
 \\
 \downarrow
 \\
 C_3(\theta_3)\ \text{as an internal spurion}
 \\
 \downarrow
 \\
 SU(3)_H,\ N_f=3
 \\
 \downarrow
 \\
 \Sigma_{ij}\sim\bar\chi_{Rj}\chi_{Li},\qquad
 \det\Sigma\supset\Sigma_{12}\Sigma_{23}\Sigma_{31}
 \\
 \downarrow
 \\
 \langle\Sigma\rangle=v_\Sigma C_3(\theta_3),\qquad
 \Lambda_{\rm 3B}\sim\Lambda_H
 \\
 \downarrow
 \\
 \kappa_{H3}(\theta_3)
 \longrightarrow
 v_{\rm EW}.
 \end{array}
 \label{eq:core_chain}
\end{equation}

The determinant interaction is a particularly useful bridge.  It is not added
solely to reproduce the desired phase: for three flavors it is a standard
cubic invariant of the composite matrix and necessarily contains the oriented
three-link product.  What remains model dependent is the vacuum alignment.
A conventional flavor-symmetric condensate would point along the identity and
would not itself produce off-diagonal link expectation values.  The
$C_3(\theta_3)$ spurion therefore carries genuine information from the
three-body sector and aligns the otherwise conventional strong dynamics.

This also clarifies the sense in which the number three is selected.  The
model does not prove a universal energetic principle that forbids four or more
sites.  Instead, three is selected first by structural minimality: it is the
smallest faithful cyclic realization of the genuine three-body source.  The
same number then reappears, rather than being independently imposed, in the
three-site internal space, the $N_f=3$ composite sector, and the cubic
determinant interaction.  In this restricted but concrete sense, the model
implements the statement that ``nature chooses three.''  A four-flavor theory
would possess a quartic determinant and naturally belong to a different
higher-body extension.  The repeated appearance of three is therefore an
organizing principle of the construction, although not yet a proof that every
occurrence of three in particle physics shares the same microscopic origin.

The parameter $\theta_3$ deserves a separate comment.  If it were promoted to
a freely dynamical angular field, the single determinant term would minimize
it at $\theta_3=-\delta_A$ modulo $2\pi$.  The present hierarchy model instead
treats $\theta_3$ as a fixed holonomy-sector parameter inherited from the
ultraviolet three-body structure.  A theory in which $\theta_3$ is itself
dynamical would require additional competing periodic terms if a nontrivial
vacuum phase were desired.

The hidden baryon and trinification are secondary extensions.  A stable
hidden baryon may furnish a composite dark-matter candidate, while the
emergent threefold organization may admit a later
$SU(3)\rightarrow SU(2)\times U(1)$ or trinification-like embedding.
Neither possibility is required for the dynamical electroweak-scale selection
mechanism developed here.

The principal quantitative limitation is nonperturbative.  The framework now
specifies a consistent EFT and its vacuum conditions, but it does not calculate
$c_A$, $m_\Sigma^2$, $\mu_A^2$, the composite pole masses, or the portal
matching from the underlying hidden gauge theory.  Those quantities determine
whether the illustrative parameter choices can be realized in a specific
microscopic theory.

\section{Conclusion}

We have constructed a four-dimensional composite toy realization of a
three-body holonomy motivated by a relativistic $(1+1)$-dimensional Dirac
system.  The low-dimensional winding is not continued literally.  Instead,
its cyclic information is encoded in a finite three-state internal sector and
represented by the spurion $C_3(\theta_3)$.

The explicit reference model is an asymptotically free hidden $SU(3)_H$ gauge
theory with three Standard-Model-singlet Dirac flavors.  Its composite matrix
$\Sigma_{ij}\sim\bar\chi_{Rj}\chi_{Li}$ contains a determinant interaction
that is cubic precisely because $N_f=3$ and includes the oriented product
$\Sigma_{12}\Sigma_{23}\Sigma_{31}$.  Hidden strong dynamics generates a
nonzero composite condensate and fixes its overall magnitude, whereas the
inherited three-body holonomy specifies its cyclic internal orientation.  With
the cyclic alignment term included, the vacuum
\begin{equation}
 \langle\Sigma\rangle=v_\Sigma C_3(\theta_3)
\end{equation}
is locally stable under the conditions derived in
Sec.~\ref{sec:cyclic_vacuum}, and
\begin{equation}
 \det\langle\Sigma\rangle
 =v_\Sigma^3e^{i\theta_3}
\end{equation}
produces the holonomy energy
\begin{equation}
 V_{\rm hol}
 =-\Lambda_{\rm 3B}^4
 \cos(\theta_3+\delta_A).
\end{equation}
Here $\delta_A$ is the independent phase carried by the anomalous determinant
coupling; it is not identified with the inherited holonomy angle $\theta_3$.

The radial composite mode is the projection along the cyclic direction,
not $\Re\Tr\Sigma$:
\begin{equation}
 S_3
 =\sqrt{\frac23}
 \Re\Tr[C_3^\dagger\Sigma].
\end{equation}
Dimensional transmutation generates $\Lambda_H$, and the strong-sector scaling
$|c_A|\sim v_\Sigma\sim\Lambda_H$ gives
\begin{equation}
 \Lambda_{\rm 3B}\sim\Lambda_H.
\end{equation}
Thus the physical three-body scale is generated dynamically rather than
inserted as an elementary scalar mass.

The coupling to the electroweak sector is most transparently described by
starting from the ordinary Higgs potential Eq.~(\ref{eq:bare_Higgs_potential})
and then adding the holonomy-dependent portal.  The leading CP-even dependence
is Eq.~(\ref{eq:kappa_theta}),
so that after the composite field condenses the Higgs quadratic coefficient is
shifted to
\begin{equation}
 m_{H,{\rm eff}}^2(\theta_3)
 =m_{H,0}^2
 +\frac12\kappa_{H3}(\theta_3)
 \langle S_3\rangle^2.
\end{equation}
Electroweak symmetry breaking occurs when
\begin{equation}
 m_{H,{\rm eff}}^2(\theta_3)<0.
\end{equation}
With $\langle H\rangle=(0,v_{\rm EW}/\sqrt2)^T$, the resulting electroweak
scale satisfies
\begin{equation}
 v_{\rm EW}^2 =-\frac{1}{\lambda_H}
 \left[
 m_{H,0}^2
 +\frac12\kappa_{H3}(\theta_3)
 \langle S_3\rangle^2
 \right].
\end{equation}
In the classically scale-invariant reference limit $m_{H,0}^2=0$, together
with $\langle S_3\rangle=c_S\Lambda_H$, this reduces to
\begin{equation}
 v_{\rm EW}
 =c_S\Lambda_{\rm match}
 \exp\left[-\frac{8\pi^2}
 {9g_H^2(\Lambda_{\rm match})}\right]
 \sqrt{-\frac{\kappa_0+\kappa_1\cos\theta_3}
 {2\lambda_H}}.
\end{equation}
The portal is therefore not the Higgs potential itself. Rather, it provides a
dynamical shift of the ordinary Higgs mass term.  The final mechanism separates
conceptually into an orientation problem and a scale-generation problem: the
three-body holonomy determines the cyclic internal orientation, hidden strong
dynamics generates the composite scale, and the portal transmits that scale to
the Higgs sector.

The present construction should not be interpreted as a complete solution to
the ultraviolet hierarchy problem.  Its more limited result is that an
electroweak-scale vacuum can be selected dynamically without introducing an
elementary mass ratio of order $10^{-17}$ by hand.  The large separation of
scales is instead encoded through dimensional transmutation driven by a
moderate hidden-sector coupling, followed by a holonomy-dependent portal to
the Higgs sector.  Possible ultraviolet threshold corrections to an elementary
Higgs remain dependent on the ultraviolet completion and are not eliminated by
the present toy framework.  A representative one-loop Coleman--Weinberg
estimate nevertheless shows that the electroweak minimum persists within the
electroweak-scale region, although a fully RG-improved analysis is required for
a quantitative statement of radiative stability.

A central message of the construction is the repeated emergence of the number
three.  The hidden $SU(3)_H$ theory is not taken to be the origin of this
number.  Rather, three first appears in the genuine three-body source, becomes
the minimal cyclic internal space, reappears in the three-flavor composite
realization, and is matched by the cubic determinant invariant.  We summarize
this organizing principle as
\begin{align}
 &\text{genuine three-body holonomy} \nonumber \\
 & \qquad \longrightarrow
 \text{minimal threefold internal organization}.
\end{align}
In this restricted sense, the phrase ``nature chooses three'' denotes a
minimality hypothesis. Three is the smallest nontrivial cyclic structure that
faithfully carries the genuine three-body information, whereas larger
multiplicities represent nonminimal extensions.  The previously considered
$SU(3)\rightarrow SU(2)\times U(1)$ breaking pattern and possible
$[SU(3)]^3$ trinification embedding can then be viewed as further developments
of the same threefold organization rather than as independent explanations of
why the number three appears.

The construction remains a proof of concept.  The map from the original
$(1+1)$-dimensional holonomy to the four-dimensional spurion is assumed, and
the composite matching coefficients are not calculated nonperturbatively.
The next decisive tests are a controlled determination of the hidden
$N_H=N_f=3$ effective potential and spectrum, and an RG-improved analysis of
the Higgs--composite effective potential.  In particular, it remains to
determine whether the required cyclic alignment and portal matching arise with
viable nonperturbative coefficients and whether the electroweak minimum
remains quantitatively stable beyond the representative one-loop estimate.
Dark matter and trinification are possible by-products, but they are not
needed for the central dynamical scale-selection mechanism.

\section{Acknowledgments}

During the preparation of this manuscript, as a non-native English speaker,
the author used ChatGPT (OpenAI GPT-5.6 Sol) for language editing, improving clarity 
and organization, and assisting with the presentation of the manuscript. 
All scientific arguments, calculations, interpretations, and conclusions 
were reviewed and verified by the author, who takes full responsibility 
for the content of the manuscript.


\begin{thebibliography}{99}

\bibitem{ColemanWeinberg1973}
S.~R.~Coleman and E.~J.~Weinberg,
``Radiative Corrections as the Origin of Spontaneous Symmetry Breaking,''
Phys.\ Rev.\ D \textbf{7}, 1888--1910 (1973).

\bibitem{GildenerWeinberg1976}
E.~Gildener and S.~Weinberg,
``Symmetry Breaking and Scalar Bosons,''
Phys.\ Rev.\ D \textbf{13}, 3333--3341 (1976).

\bibitem{Farzinnia2013}
A.~Farzinnia, H.-J.~He, and J.~Ren,
``Natural Electroweak Symmetry Breaking from Scale Invariant Higgs Mechanism,''
Phys.\ Lett.\ B \textbf{727}, 141--150 (2013).
arXiv:1308.0295 [hep-ph].

\bibitem{Chway2014}
D.~Chway, R.~Derm\'i\v{s}ek, T.~H.~Jung, and H.~D.~Kim,
``Radiative Electroweak Symmetry Breaking Model Perturbative
All the Way to the Planck Scale,''
Phys.\ Rev.\ Lett.\ \textbf{113}, 051801 (2014).

\bibitem{Lebedev2021}
O.~Lebedev,
``The Higgs Portal to Cosmology,''
Prog.\ Part.\ Nucl.\ Phys.\ \textbf{120}, 103881 (2021).

\bibitem{Munakata1990}
Y.~Munakata, J.~Sakamoto, T.~Ino, T.~Nakamae, and F.~Yamamoto,
``Exact Solutions of a Relativistic Three-Body Bound State in $1+1$ Dimensions,''
Prog.\ Theor.\ Phys.\ \textbf{83}, 84--96 (1990).

\bibitem{Sakamoto1993}
J.~Sakamoto,
``Construction of $N$ Body Bound State Solution of the Dirac Particles
in $1+1$ Dimensional Space-Time,''
Prog.\ Theor.\ Phys.\ \textbf{89}, 119--130 (1993).

\bibitem{SakamotoNakanoYoshikawa1993}
J.~Sakamoto, T.~Nakano, and T.~Yoshikawa,
``Two Types of N-Body Bound State Solutions of the Dirac Particles with the $\delta$ Function Potential 
in 1+1 Dimensional Space-Time,''
Prog. Theor. Phys. \textbf{89}, 767--770 (1993).

\bibitem{Yoshikawa2026phase}
T.~Yoshikawa,
``Phase Encoding of Genuine Three-Body Interactions in a Relativistic
Dirac System in $1+1$ Dimensions,''
arXiv:2609.25478 (2026).

\bibitem{Yoshikawa2026family}
T.~Yoshikawa,
``Three-Body Holonomy as a Toy-Model Mechanism for Family Triplication
and Mass Hierarchy in $(1+1)$ Dimensions,''
arXiv:2609.24022 (2026).

\bibitem{Yoshikawa2026SU3}
T.~Yoshikawa,
``Three-Body Holonomy and $SU(3)\to SU(2)\times U(1)$ Symmetry Breaking,''
arXiv:2610.00235 (2026).

\bibitem{NelsonParkWalker2019}
A.~E.~Nelson, M.~Park, and D.~G.~E.~Walker,
``Composite Higgs Models with a Hidden Sector,''
Phys.\ Rev.\ D \textbf{100}, 076015 (2019).

\bibitem{GrossWilczek1973}
D.~J.~Gross and F.~Wilczek,
``Ultraviolet Behavior of Non-Abelian Gauge Theories,''
Phys.\ Rev.\ Lett.\ \textbf{30}, 1343--1346 (1973).

\bibitem{Politzer1973}
H.~D.~Politzer,
``Reliable Perturbative Results for Strong Interactions?,''
Phys.\ Rev.\ Lett.\ \textbf{30}, 1346--1349 (1973).

\bibitem{tHooft1976}
G.~'t~Hooft,
``Computation of the Quantum Effects Due to a Four-Dimensional Pseudoparticle,''
Phys.\ Rev.\ D \textbf{14}, 3432--3450 (1976);
Erratum Phys.\ Rev.\ D \textbf{18}, 2199 (1978).

\bibitem{PisarskiWilczek1984}
R.~D.~Pisarski and F.~Wilczek,
``Remarks on the Chiral Phase Transition in Chromodynamics,''
Phys.\ Rev.\ D \textbf{29}, 338--341 (1984).

\bibitem{BabuHePakvasa1986}
K.~S.~Babu, X.-G.~He, and S.~Pakvasa,
``Neutrino Masses and Proton Decay Modes in
$SU(3)\times SU(3)\times SU(3)$ Trinification,''
Phys.\ Rev.\ D \textbf{33}, 763--772 (1986).

\bibitem{Stech2014}
B.~Stech,
``Trinification Phenomenology and the Structure of Higgs Bosons,''
JHEP \textbf{08}, 139 (2014).
arXiv:1403.2714 [hep-ph].

\end{thebibliography}
\end{document}